\documentclass[11pt,a4paper]{article}

\usepackage[T1]{fontenc}
\usepackage[utf8]{inputenc}
\usepackage[english]{babel}

\usepackage[numbers,sort&compress]{natbib}

\usepackage[a4paper,margin=2.5cm]{geometry}

\usepackage{lmodern}

\usepackage{indentfirst}
\usepackage{amsmath,amssymb,amsfonts}
\usepackage{bm}
\usepackage{braket}
\usepackage{cancel}
\usepackage{relsize}
\usepackage{graphicx}
\usepackage{overpic}
\usepackage{float}
\usepackage{tikz}

\usepackage[colorlinks=true,
            linkcolor=blue,
            citecolor=blue,
            urlcolor=blue]{hyperref}
\usepackage{doi}  % renders DOI fields in the bibliography as clickable links

\title{Covariant extrinsic curvature expansion of the nonlocal effective action for a massless scalar field on a manifold with boundary}

\author{
A. Boasso, C.~D.~Fosco, B.~C.~Guntsche, and F.~D.~Mazzitelli\\[0.5em]
{\normalsize\it Centro At\'omico Bariloche and Instituto Balseiro}\\
{\normalsize\it Comisi\'on Nacional de Energ\'ia At\'omica}\\
{\normalsize\it R8402AGP Bariloche, Argentina}
}

\date{\today}

\begin{document}

\maketitle

%====================================================================
\begin{abstract}
We study the nonlocal effective action of a massless scalar field defined on a flat manifold with a curved boundary. Using a heat-kernel approach, we derive a covariant expansion of the nonlocal contribution to quadratic order in the extrinsic curvature tensor. Our
construction provides a geometric framework that both reproduces earlier results
obtained for Monge-patch embeddings and extends them to more general surfaces
that need not admit a global Monge-patch description. The expansion is valid in
the regime where gradients of the extrinsic curvature dominate over nonlinear
curvature effects. As an application, we compute the particle-creation rate for
an oscillating deformed ring in $2+1$ dimensions and an oscillating deformed
sphere in $3+1$ dimensions.
\end{abstract}

%====================================================================
\section{Introduction}\label{sec:intro}

The one-loop effective action of a quantum field encodes, in a single
functional, two qualitatively different kinds of information. Its analytic,
local part controls the ultraviolet divergences and is absorbed into the
renormalisation of the bare couplings of the theory. Its nonanalytic, nonlocal
part is finite, calculable, and physically responsible for genuinely quantum
phenomena: vacuum polarisation, conformal anomalies, Hawking radiation, the
Casimir force, particle creation by time-dependent backgrounds, and dissipative
backreaction on classical sources
\cite{birrell1982,Parker2009,Wald1995,Buchbinder2021,Calzetta2008,DEWITT1975}.
While the local sector is by now textbook material, controlled covariantly
through the heat-kernel or DeWitt-Schwinger expansion
\cite{Vassilevich,Kirsten,avramidi1991}, the nonlocal sector is harder to
extract because it is, by definition, invisible to any finite-order expansion
in derivatives.

A useful strategy to access the nonlocal sector while remaining within the
heat-kernel framework was developed by Vilkovisky and collaborators, and later
by Barvinsky and Vilkovisky
\cite{Vilkovisky:1984,Barvinsky1987,Barvinsky1990,Barvinsky1994,Barvinsky:1994hw,vilkovisky,quantum_noise}.
Rather than truncating the heat-kernel expansion at a fixed order in proper
time, one truncates at a fixed order in some background invariant (typically
curvature). The local divergent
coefficients of the heat-kernel expansion, organised in this way, then fix (by
analyticity) the full nonlocal kernel that multiplies the chosen invariant.
The same logic was later refined and applied in a variety of contexts: running
of Newton's constant and noninteger powers of the d'Alembertian
\cite{FDM-resumation,Dalvit:1994,lopez_nacir_fdm}, the conformal
anomaly action \cite{Riegert:1984,Mottola:2006,Barvinsky2023}, nonlocal
cosmology and gravitational radiation
\cite{Donoghue1993,Donoghue1994,Donoghue:2014,calmet_grav,Calmet2019,Calmet2021},
and most recently the construction of effective actions in $D$ dimensions and in
nontrivial cosmological and astrophysical backgrounds
\cite{nosotros_cotton,bardeen,Shapiro:2008sf}.

Once the field lives on a manifold $\mathcal{M}$ with boundary $\partial\mathcal{M}$, the heat-kernel expansion
picks up an additional set of \emph{boundary} contributions, with coefficients
that encode the local boundary condition and depend only on the extrinsic
geometry of $\partial\mathcal{M}$ and its covariant derivatives
\cite{Vassilevich,Branson:1995,Kirsten}. The corresponding effective action
governs much of the physics of mirrors, films, and dielectrics. Its real part
generates the static Casimir energy of arbitrary geometries
\cite{BordagBook,Plunien1986,Bordag2009,Milton2001}; its imaginary part, when
the boundary is allowed to move, generates the dynamical Casimir effect,
the spontaneous emission of correlated pairs of quanta out of the vacuum by
an accelerated mirror, conjectured by Moore \cite{Moore1970} and computed
explicitly by Davies and Fulling for one-dimensional mirrors
\cite{DaviesFulling1976,DaviesFulling1977}. The dynamical Casimir effect has
since been studied for a wide range of geometries, boundary conditions, and
quantum statistics, and reviewed extensively
\cite{Dodonov2010,Dalvit:2010,Nation2012,Dodonov2020}.

Two of us recently considered precisely this dynamical setting, working in a
flat $(d{+}1)$-dimensional bulk with a single moving boundary supporting 
Dirichlet \cite{Bruno:Dirichlet} or Neumann \cite{Bruno:Neumann} conditions. 
The boundary in those works was parametrised as a Monge patch, that is, as a global graph
\(x_{d+1}=\psi(y^a)\) of a height function over a hyperplane. Within that
parametrisation, the nonlocal effective action could be reduced to a quadratic
form in $\psi$ whose kernel, for even $d$, contains a logarithm of the
boundary Laplacian and, for odd $d$, a fractional power of it: exactly the
nonanalytic structure responsible for the imaginary part and the radiation rate.

The Monge-patch parametrisation, however, cannot globally describe systems with
closed boundaries (cylinders, spheres, tori), nor surfaces
with overhang, multivalued sections, or non-trivial topology; and even when
it does apply, the result, written in terms of $\psi$ and flat Cartesian
derivatives, conceals the geometric content of the answer. In the present
work we lift the analysis to a manifestly covariant, geometric formulation in
which the natural variable is the extrinsic curvature tensor $K_{ab}$ of
$\partial\mathcal{M}$ and the natural building blocks are its
diffeomorphism-invariant scalars. Using the standard boundary heat-kernel
coefficients of Branson, Gilkey and Vassilevich
\cite{Vassilevich,Branson:1995} together with the reconstruction procedure 
\cite{Vilkovisky:1984,FDM-resumation}, we derive the unique nonlocal 
contribution to the effective
action at quadratic order in $K_{ab}$, valid in the regime in which gradients
of $K_{ab}$ dominate over higher powers of $K_{ab}$ itself,
\(\nabla\nabla K \gg K^{3}\),  in close analogy with the Barvinsky-Vilkovisky
expansion of the bulk effective action in semiclassical gravity. We show that
the Monge-patch results \cite{Bruno:Dirichlet,Bruno:Neumann} are recovered as
a special case, providing both a nontrivial cross-check and, by comparison,
the determination of the overall coefficient $\kappa_d$ in odd $d$, a
constant which the heat-kernel coefficients alone are not able to fix.

The covariant formulation has direct physical consequences. For an arbitrary
oscillating boundary, the extrinsic curvature appears bilinearly in the action
through a kernel of the form $(-\nabla^2)^{d/2-1}\log(-\nabla^2)$ in even $d$
and $(-\nabla^2)^{d/2-1}$ in odd $d$, where $\nabla^2$ is the Laplace-Beltrami
operator of the worldtube $\partial\mathcal{M}$ itself, regardless of whether
that worldtube admits a global Monge description. As applications, we compute
the particle-creation rate for an oscillating deformed cylinder
(``pulsating ring'') in $2{+}1$ dimensions and for an oscillating multipolar
deformation of a sphere in $3{+}1$ dimensions. The latter exhibits a clean
mode-by-mode threshold structure: the multipole $\ell$ radiates only above a
characteristic gap frequency $\omega_\ell=\sqrt{\ell(\ell{+}1)}/R$ set by the
spherical-harmonic eigenvalue, and the rates for Dirichlet and Neumann
boundary conditions differ by a fixed numerical ratio.

The paper is organised as follows. In Section~\ref{sec:heat-kernel} we collect
the heat-kernel ingredients and introduce our geometric conventions. In
Section~\ref{sec:nonlocal} we derive the nonlocal effective action at quadratic
order in $K_{ab}$, separately for even and odd boundary dimension. In
Section~\ref{sec:monge} we make contact with the Monge-patch results of
\cite{Bruno:Dirichlet,Bruno:Neumann} and use them to fix the overall coefficient
of the odd-$d$ result. In Section~\ref{examples} we work out two examples in
detail: the oscillating ring (Section~\ref{sec:ring}) and the oscillating
multipolar sphere (Section~\ref{sec:multipole}). Section~\ref{sec:conclusions}
contains our conclusions; Appendix~\ref{app:bogoliubov} relates
$\mathrm{Im}\,\Gamma$ to the Bogoliubov-coefficient computation of
\cite{sphere} and provides a consistency check of our formulas.

%====================================================================
\section{Heat kernel on manifolds with boundary}\label{sec:heat-kernel}

In this section we review some basic aspects of the heat-kernel expansion of the
effective action and introduce the geometric notions we will use. We consider a
free real scalar field $\phi$ with mass $m$ on a $(d+1)$-dimensional manifold
$\mathcal{M}$. From now on, we will refer to $\mathcal{M}$ as the bulk, and we
assume that $\mathcal{M}$ is a Riemannian manifold, i.e., with Euclidean signature. Although our main interest
will be the massless case, we keep $m$ at this stage, where it can also be
regarded as an infrared regulator. The equation of motion for the scalar field
in the bulk is
\begin{equation}
\underbrace{(-\nabla^{\mu}\nabla_{\mu}+m^2)}_{:=\widehat{\mathcal O}}\,\phi=\left(-\frac{1}{\sqrt{g}}\partial_{\mu}\left(\sqrt{g}\,g^{\mu\nu}\partial_{\nu}\right)+m^2\right)\phi=0\,,
    \label{eom}
\end{equation}
where $g_{\mu\nu}=g_{\mu\nu}(x)$ is the metric tensor on $\mathcal{M}$, $g=\det g_{\mu\nu}$, and $\nabla_{\mu}$ is the associated covariant derivative.

We also assume that the bulk has a smooth, nonempty boundary
$\partial\mathcal{M}$. This boundary is a codimension-one submanifold of
$\mathcal{M}$, so that $\dim\partial\mathcal{M}=d$. Moreover, we take
$\partial(\partial\mathcal{M})=\emptyset$. On the boundary, the scalar field is
required to satisfy one of the following local boundary conditions:
\begin{center}
\begin{tabular}{r l}
 Dirichlet boundary condition (D): &\qquad $\phi|_{\partial\mathcal{M}}(y)=0$\,, \\
 Neumann boundary condition (N): &\qquad $n^{\mu}\nabla_{\mu}\phi|_{\partial\mathcal{M}}(y)=0$\,,
\end{tabular}
\end{center}
for all $y\in\partial\mathcal{M}$, where $n^{\mu}$ is the outward-pointing unit
normal to $\partial\mathcal{M}$. We use $x^{\mu}$, with $\mu=1,\dots,d+1$, to
denote coordinates on the bulk, and $y^{a}$, with $a=1,\dots,d$, to denote
coordinates on the boundary. Greek indices refer to bulk coordinates, whereas
Latin indices refer to boundary coordinates.

Within the path-integral formalism, the one-loop effective action $\Gamma$ of the theory can be written as
\begin{equation}
  \Gamma^{(\mathrm{D},\mathrm{N})}=\frac{1}{2}\mathrm{Tr}\left[\log \widehat{\mathcal O}^{(\mathrm{D},\mathrm{N})}\right]\,,
\end{equation}
where the boundary conditions are encoded in the domain of the operator $\widehat{\mathcal O}$ defined in Eq.\eqref{eom}.
Equivalently, the trace is taken over the modes satisfying the chosen boundary
condition. In what follows, we will omit the superscripts referring to the
boundary conditions when no confusion can arise. The logarithm of the operator
$\widehat{\mathcal O}$ can also be expressed formally as an integral over proper time,
\begin{equation}
  \log \widehat{\mathcal O} = -\int_{0}^{\infty}\frac{\mathrm{d}s}{s}\,e^{-s\,\widehat{\mathcal O}}\,,
\end{equation}
up to an additive constant independent of $\widehat{\mathcal O}$. Therefore,
\begin{equation}
  \Gamma=\frac{1}{2}\mathrm{Tr}\left[\log (-\nabla^{\mu}\nabla_{\mu}+m^2)\right] = -\frac{1}{2}\int_{0}^{\infty}\frac{\mathrm{d}s}{s}e^{-sm^2}\mathrm{Tr}\left[e^{s\nabla^{\mu}\nabla_{\mu}}\right]\,.
  \label{eff_action0}
\end{equation}
The trace in Eq.~\eqref{eff_action0} is the heat trace of the operator
$-\nabla^\mu\nabla_\mu$. It can be computed from the coincidence limit of the
heat kernel $\langle x|e^{s\nabla^{\mu}\nabla_{\mu}}|x'\rangle$
\cite{Vassilevich}, which satisfies the heat equation
\begin{equation}
  \left(\partial_s-\nabla^{\mu}\nabla_{\mu}\right)\langle x|e^{s\nabla^{\mu}\nabla_{\mu}}|x'\rangle=0\,,
  \label{heat_kernel}
\end{equation}
where the covariant derivatives act on the coordinate $x$, with the initial condition
\begin{equation}
    \lim_{s\to 0^+}\langle x|e^{s\nabla^{\mu}\nabla_{\mu}}|x'\rangle=\frac{1}{\sqrt{g(x)}}\delta^{d+1}(x-x')\,.
\end{equation}
For $s\to0^{+}$, the heat trace admits the asymptotic expansion \cite{Vassilevich,Kirsten,birrell1982}
\begin{equation}
  \mathrm{Tr}\left[e^{s\nabla^{\mu}\nabla_{\mu}}\right]\approx(4\pi s)^{-\frac{d+1}{2}}\sum_{n=0}^{\infty}s^{n/2}\left(\int_{\mathcal{M}}a_{n/2}(x)+\sqrt{4\pi}\int_{\partial\mathcal{M}}b_{n/2}(y)\right)\,,
  \label{kernel1}
\end{equation}
where the symbol $\approx$ denotes equality in the sense of a short-proper-time
asymptotic expansion. The quantities $a_{n/2}(x)$ and $b_{n/2}(y)$, with
$n\in\mathbb{N}_0$, are the heat-kernel, or DeWitt-Schwinger, coefficients. The
integral on $\mathcal{M}$ implicitly has the volume element of the bulk,
$\sqrt{g}\,\mathrm{d}^{d+1}x$, while the integral on $\partial\mathcal{M}$ has
the volume element of the boundary, $\sqrt{h}\,\mathrm{d}^{d}y$, with $h=\det
h_{ab}$, where $h_{ab}$ is the metric induced on $\partial\mathcal{M}$. The
coefficients can be obtained, for example, by substituting the expansion
Eq.~\eqref{kernel1} into the heat equation in Eq.~\eqref{heat_kernel} before
taking $x=x'$, together with the corresponding boundary conditions, but we do
not pursue that route here.

By inserting the short-proper-time expansion in Eq.~\eqref{kernel1} into
Eq.~\eqref{eff_action0}, and exchanging the sum with the integrals at the formal
level, we obtain the heat-kernel contribution to the effective action,
\begin{multline}
  \Gamma^{\mathrm{UV}}=-\frac{(4\pi)^{-\frac{d+1}{2}}}{2}\int_{\mathcal{M}}\int_{0}^{\infty}\mathrm{d}s\,s^{-\frac{d+3}{2}}e^{-sm^2}\left(\sum_{n=0}^{\infty}s^{\frac{n}{2}}a_{n/2}(x)\right) \\
  -\frac{(4\pi)^{-\frac{d}{2}}}{2}\int_{\partial\mathcal{M}}\int_{0}^{\infty}\mathrm{d}s\,s^{-\frac{d+3}{2}}e^{-sm^2}\left(\sum_{n=0}^{\infty}s^{\frac{n}{2}}b_{n/2}(y)\right)\,.
  \label{eff_action_complete}
\end{multline}
Equation~\eqref{eff_action_complete} should not be understood as an exact
expression for the full effective action over the whole range of proper time.
Rather, it represents the contribution generated by the short-proper-time
expansion, which controls the UV divergences and will be used in the next
section to reconstruct the corresponding nonlocal terms. The coefficients
$a_{n/2}$ and $b_{n/2}$ are local geometric quantities and do not depend on $s$
or $m$. Within this formalism, for the boundary conditions considered here, the
dependence on the boundary condition is encoded in the boundary coefficients
$b_{n/2}$. These coefficients are local scalar quantities constructed from the
geometry of $\partial\mathcal{M}$ and differ for Dirichlet and Neumann boundary
conditions.

In what follows, we take the bulk to be flat. Then the relevant boundary
invariants can be written in terms of the extrinsic curvature tensor and its
covariant derivatives. We define the extrinsic curvature as
\begin{equation}
    K_{ab}=\frac{\partial x^{\mu}}{\partial y^a}\frac{\partial x^{\nu}}{\partial y^b}\nabla_{\mu}n_{\nu}\,,
\end{equation}
where $x^\mu=x^\mu(y)$ is the embedding of the boundary in the bulk. Latin
indices are raised and lowered with the induced metric on $\partial\mathcal{M}$,
and $\nabla_a$ denotes the covariant derivative compatible with that metric. We
also define the mean curvature $K$ as the trace of the extrinsic curvature
tensor:
\begin{equation}
    K=h^{ab}K_{ab}\,.
\end{equation}

For both Dirichlet and Neumann boundary conditions, the bulk coefficients of the
heat-kernel expansion are
\begin{equation}
  a_0^{(\mathrm{D},\mathrm{N})}=1,\qquad
  a_{n/2}^{(\mathrm{D},\mathrm{N})}=0\,,\qquad n\geq 1\,,
\end{equation}
while the first boundary coefficients are, for the Dirichlet boundary condition
\cite{Vassilevich,Branson:1995},
\begin{equation}
  \begin{aligned}
    b_0^{(\mathrm{D})} &= 0\,,\\
    b_{1/2}^{(\mathrm{D})} &= -\frac{1}{4}\,,\\
    b_1^{(\mathrm{D})} &= \frac{1}{3\sqrt{4\pi}}K\,,\\
    b_{3/2}^{(\mathrm{D})} &= \frac{1}{384}\left(10K_{ab}K^{ab}-7K^2\right)\,,\\
    b_{2}^{(\mathrm{D})} &= \frac{1}{360\sqrt{4\pi}}\left[24\nabla^a\nabla_a K +\frac{1}{21}\left(40K^3-264KK^{ab}K_{ab}+320K_{a}{}^{b}K_{b}{}^{c}K_{c}{}^{a}\right)\right]\,,\\
    b_{5/2}^{(\mathrm{D})} &= -\frac{1}{2048}\nabla^cK^{ab}\nabla_cK_{ab}+\mathcal{O}(K^4)\,.
  \end{aligned}
  \label{coeff_dirichlet}
\end{equation}
For the Neumann boundary condition, they are
\begin{equation}
  \begin{aligned}
    b_0^{(\mathrm{N})} &= 0\,,\\
    b_{1/2}^{(\mathrm{N})} &= \frac{1}{4}\,,\\
    b_1^{(\mathrm{N})} &= \frac{1}{3\sqrt{4\pi}}K\,,\\
    b_{3/2}^{(\mathrm{N})} &= \frac{1}{384}\left(2K_{ab}K^{ab}+13K^2\right)\,,\\
    b_{2}^{(\mathrm{N})} &= \frac{1}{360\sqrt{4\pi}}\left[24\nabla^a\nabla_a K +\frac{1}{21}\left(280K^3-168KK^{ab}K_{ab}+224K_{a}{}^{b}K_{b}{}^{c}K_{c}{}^{a}\right)\right]\,,\\
    b_{5/2}^{(\mathrm{N})} &= -\frac{19}{2048}\nabla^cK^{ab}\nabla_cK_{ab}+\mathcal{O}(K^4)\,.
  \end{aligned}
  \label{coeff_neumann}
\end{equation}

%====================================================================
\section{Nonlocal effective action for a massless theory}\label{sec:nonlocal}

Our goal is to obtain the nonlocal part of the effective action in
Eq.~\eqref{eff_action_complete} for a massless scalar field in a flat bulk, with
either Dirichlet or Neumann boundary conditions. The nonlocal part of the
effective action contains the contributions that cannot be expressed in terms of
analytic functions of covariant derivatives. It therefore encodes the
nonanalyticities of the action. In particular, after analytic continuation to
the corresponding Lorentzian theory, the imaginary part of the effective action
is contained in this nonlocal piece, making it relevant for the study of
particle creation.

To this end, we employ a procedure introduced in previous works
\cite{Vilkovisky:1984,FDM-resumation}, in which the nonlocal part of the
effective action is reconstructed from the divergences associated with a finite
number of heat-kernel coefficients. In those works, the method was applied to
curved bulks without boundary, and the resulting actions correspond to
semiclassical gravity theories. Here, instead, we apply this technique to a
theory in a flat bulk with a nontrivial boundary. The results obtained in this
section can be used, for example, to study particle creation induced by the
dynamical Casimir effect for a broad class of possible boundaries.

As mentioned above, the key idea in Refs.~\cite{Vilkovisky:1984,FDM-resumation}
is to obtain the nonlocal part of the effective action using only the
information contained in the divergent coefficients of the heat-kernel
expansion. Let us focus on the boundary coefficients $b_{n/2}$ in
Eq.~\eqref{eff_action_complete} and perform a dimensional analysis before
proceeding. In units where energy has dimension $+1$, the relevant dimensions
are \begin{equation} [\nabla_a]=[K_{ab}]=1\,,\qquad [b_{n/2}]=n-1\,.
\end{equation} Since each coefficient $b_{n/2}$ is a local scalar constructed
from covariant geometric quantities, at fixed $n$ it may contain $K$, $K_{ab}$,
covariant derivatives, and their contractions, with total dimension $n-1$. No
additional geometric objects, such as the normal vector to
$\partial\mathcal{M}$, appear explicitly in the $b_{n/2}$'s, since any
occurrence of them can be rewritten in terms of $K_{ab}$ and covariant
derivatives on $\partial\mathcal{M}$. In a flat bulk, the local geometric
invariants of a smooth codimension-one boundary can therefore be written in
terms of $K_{ab}$ and its covariant derivatives.

To compute the effective action, one would have to sum the series in
Eq.~\eqref{eff_action_complete}; however, in practice, the full series is not
available. To proceed, instead of applying a direct truncation in the
heat-kernel order, we organise the expansion according to the number of powers
of the extrinsic curvature.\footnote{From now on, by powers of the extrinsic
curvature we mean powers of $K$ and/or $K_{ab}$.} Let us analyse this point in
more detail.

Zeroth-order terms provide only contributions proportional to the volumes of
$\mathcal{M}$ and $\partial\mathcal{M}$, and they are absorbed into the bare
constants of the theory through renormalization. At first order in the extrinsic
curvature, one obtains either $K$ itself, integrated over $\partial\mathcal{M}$,
which gives a local contribution to the effective action, or terms such as
$\nabla^a\nabla_a\cdots\nabla^b\nabla_bK$ and
$\nabla^a\nabla_a\cdots\nabla^b\nabla^cK_{bc}$, which are total derivatives and
do not contribute after integration over $\partial\mathcal{M}$, since
$\partial(\partial\mathcal{M})=\emptyset$. We therefore focus on the quadratic
contributions in the extrinsic curvature, which give the first nontrivial
contribution to the nonlocal part of the effective action.

It is useful to note that the integer-indexed coefficients $b_n$ do not contain
quadratic terms in the extrinsic curvature. Indeed, before contractions with
derivatives, a quadratic term contains $0$, $2$, or $4$ free boundary indices
carried by the two extrinsic curvatures. However, dimensional analysis implies
that the quadratic contributions to the integer-indexed coefficients would
contain an odd number of derivatives, and no scalar covariant term of this form
can be constructed. Therefore, at quadratic order in the extrinsic curvature,
the effective action can be written schematically as \begin{equation}
	\Gamma^{\mathrm{UV}(2)} = -\frac{1}{2(4\pi)^{d/2}}
	\int_{0}^{\infty}\frac{\mathrm{d}s}{s^{d/2}}\,e^{-sm^2}
	\sum_{k=0}^{\infty}s^k \sum_{i}
	\int_{\partial\mathcal{M}}\mathcal{Q}_{k,i}\,, \label{eff_action_2}
\end{equation} where we have used the fact that only heat-kernel coefficients
with half-integer index, starting from $b_{3/2}$, contribute at this order. Here
$\mathcal{Q}_{k,i}$ denotes a generic scalar term quadratic in $K_{ab}$ with
$2k$ covariant derivatives, together with a real coefficient determined by the
corresponding heat-kernel coefficient $b_{k+3/2}$. For instance, from
$b_{3/2}^{(\mathrm{D})}$ in Eq.~\eqref{coeff_dirichlet}, we can identify
$\mathcal{Q}_{0,1}^{(\mathrm{D})}=\frac{10}{384}K^{ab}K_{ab}$ and
$\mathcal{Q}_{0,2}^{(\mathrm{D})}=-\frac{7}{384}K^2$, while from
$b_{5/2}^{(\mathrm{D})}$ we obtain
$\mathcal{Q}_{1,1}^{(\mathrm{D})}=-\frac{1}{2048}\nabla^{c}K^{ab}\nabla_{c}K_{ab}$.

Keeping only the terms quadratic in the extrinsic curvature amounts to working
in a regime where the curvature is perturbatively small, while its derivatives
are not necessarily treated as small. More precisely, the approximation consists
in neglecting cubic and higher powers of the extrinsic curvature in comparison
with terms containing two powers of the curvature and an arbitrary number of
covariant derivatives. Schematically, this requires \begin{equation}
\nabla\nabla K\gg K^3\,.  \end{equation} In this sense, Eq.~\eqref{eff_action_2}
represents the leading contribution to the heat-kernel effective action in an
expansion in powers of the extrinsic curvature.

Despite the apparently complicated form of Eq.~\eqref{eff_action_2}, all
quadratic terms can be reduced to two simpler structures. By repeated
integration by parts on $\partial\mathcal{M}$, and using the Codazzi identity
for a hypersurface embedded in a flat bulk, \begin{equation} \nabla_a
K_{bc}-\nabla_b K_{ac}=0\,, \label{geometry3} \end{equation} all the quadratic
structures $\mathcal{Q}_{k,i}$ can be written, under the integral over
$\partial\mathcal{M}$, in terms of \begin{equation} K(-\nabla^2)^{k}K\,,
\qquad\text{and}\qquad K^{ab}(-\nabla^2)^{k}K_{ab}\,.  \end{equation} Here
$\nabla^2\equiv\nabla^a\nabla_a$ is the Laplace-Beltrami operator on the
boundary.

Moreover, for terms with $k\geq1$, the covariant derivatives can be commuted at
the price of generating only higher-order terms in the extrinsic curvature.
Indeed, when acting on a boundary tensor, the commutator of covariant
derivatives produces the intrinsic Riemann tensor of the boundary, so that,
schematically, $[\nabla,\nabla]\sim R$. On the other hand, for a hypersurface
embedded in a flat bulk, the Gauss identity gives \begin{equation}
R_{abcd}=K_{ac}K_{bd}-K_{ad}K_{bc}\,, \end{equation} so that schematically
$[\nabla,\nabla]\sim R\sim K^2$. Therefore, when such commutators appear in
terms already quadratic in $K_{ab}$, they generate contributions of order
$\mathcal{O}(K^4)$. Thus, up to terms of order $\mathcal{O}(K^4)$ and total
derivatives, all the contributions with $k\geq1$ can then be written in terms of
a single scalar structure, \begin{equation} K(-\nabla^2)^kK\,.  \end{equation}

To complete the reduction, we have to treat separately the derivative-free terms
coming from the coefficient $b_{3/2}$, namely, $K^2$ and $K^{ab}K_{ab}$. By
taking a covariant derivative of the Codazzi identity in Eq.~\eqref{geometry3}
and contracting the corresponding indices, we obtain \begin{equation} \nabla^2
	K_{ab}=\nabla_a\nabla_b K+\mathcal{O}(K^3)\,.  \end{equation} Therefore,
	\begin{equation}
	K_{ab}=\frac{\nabla_a\nabla_b}{\nabla^2}K+\mathcal{O}(K^3)\,,
\end{equation} up to the addition of a homogeneous contribution annihilated by
$\nabla^2$. This homogeneous contribution is not fixed by the local equation,
but by the prescription chosen for the inverse of the Laplace-Beltrami
operator. In what follows, we choose this prescription so as to discard such
homogeneous modes, in analogy with the treatment of nonlocal curvature
invariants in Ref.~\cite{Barvinsky:1994hw}. Using this expression inside the
integral, and integrating by parts while keeping the same prescription for the
inverse Laplace-Beltrami operator, one obtains 
\begin{equation}
\int_{\partial\mathcal{M}}K^{ab}K_{ab} =
\int_{\partial\mathcal{M}}K^2+\mathcal{O}(K^4)\,.
\label{eq:KabKab-to-K2}
\end{equation} Therefore, all
the quadratic structures in Eq.~\eqref{eff_action_2} can be reduced to a single
quadratic form, namely, \begin{equation}
	\Gamma^{\mathrm{UV}(2)}=-\frac{1}{2(4\pi)^{d/2}}\int_{\partial\mathcal{M}}\int_{0}^{\infty}\frac{\mathrm{d}s}{s^{d/2}}\,e^{-sm^2}\left[\sum_{k=0}^{\infty}\alpha_k\,K(-\nabla^{2})^{k}
	K \,s^{k}\right]\,, \label{eff_action_2_0} \end{equation} where each
	coefficient $\alpha_k$ is obtained from the heat-kernel coefficient
	$b_{k+3/2}$, after the manipulations described above. This reduction
	allows us to read the first coefficients $\alpha_k$ directly from the
	boundary heat-kernel coefficients listed above. For instance, from
	Eq.~\eqref{coeff_dirichlet} we obtain \begin{equation} \begin{aligned}
	\alpha_0^{(\mathrm{D})}&=\frac{10}{384}-\frac{7}{384}=\frac{1}{128}\,,\\
\alpha_1^{(\mathrm{D})}&=-\frac{1}{2048}\,; \end{aligned} \label{alphaD}
\end{equation} while from Eq.~\eqref{coeff_neumann} we obtain \begin{equation}
\begin{aligned}
\alpha_0^{(\mathrm{N})}&=\frac{2}{384}+\frac{13}{384}=\frac{5}{128}\,,\\
\alpha_1^{(\mathrm{N})}&=-\frac{19}{2048}\,.  \end{aligned} \label{alphaN}
\end{equation}

The reduction of the generic form of the second-order effective action in
Eq.~\eqref{eff_action_2} to a single scalar structure in
Eq.~\eqref{eff_action_2_0} is crucial to capture the associated nonlocal
contribution. Following the procedure presented in Refs.~\cite{Vilkovisky:1984,
FDM-resumation} to extract the nonlocal contribution to the effective action, we
take the massless limit in Eq.~\eqref{eff_action_2_0} and, at least formally,
sum the series. This gives \begin{equation}
\Gamma^{\mathrm{UV}(2)}=-\frac{1}{2(4\pi)^{d/2}}\int_{\partial\mathcal{M}}\int_{0}^{\infty}\frac{\mathrm{d}s}{s^{d/2}}\Big(K\alpha(-s\nabla^{2})
K\Big)\,, \label{eff_action_2_1} \end{equation} where \begin{equation}
\alpha(\xi)=\sum_{k=0}^{\infty}\alpha_k\, \xi^k\,.  \label{eff_action_2_2}
\end{equation} In taking the limit $m\to0$, we assume only that the integral
\begin{equation} \int_{0}^{\infty}\frac{\mathrm{d}\xi}{\xi^{d/2}}\,\alpha(\xi)
\end{equation} converges at the upper limit.

To extract the nonlocal contribution, or equivalently the terms involving
nonholomorphic functions of derivatives, we do not need to know the complete
function $\alpha(\xi)$ in the even-dimensional case. Instead, we can study the
ultraviolet divergences in Eq.~\eqref{eff_action_2_1}, which arise from the
lower limit of the proper-time integral, and deduce the corresponding nonlocal
contribution from them. We therefore introduce a cutoff $\Lambda$ with
dimensions of energy and write \begin{equation}
	\int_{\frac{1}{\Lambda^2}}^{\infty}\frac{\mathrm{d}s}{s^{d/2}}\,\alpha(-s\nabla^2)
	= (-\nabla^2)^{\frac{d}{2}-1}
	\int_{\frac{-\nabla^2}{\Lambda^2}}^{\infty}
	\frac{\mathrm{d}\xi}{\xi^{d/2}}\,\alpha(\xi)\,,
\label{proper_time_cutoff} \end{equation} where we have used the fact that
$-\nabla^2$ is a non-negative operator. Possible zero modes are understood to be
treated according to the prescription specified above. We identify two possible
structures in the limit $\Lambda\to\infty$.

\subsection{Even-dimensional boundary}

For even $d$, we obtain \begin{multline}
	\int_{\frac{1}{\Lambda^2}}^{\infty}\frac{\mathrm{d}s}{s^{d/2}}\,\alpha(-s\nabla^2)
	= \text{const.}\times(-\nabla^2)^{\frac{d}{2}-1}
	-\sum_{k=0}^{\frac{d}{2}-2} \frac{2\,\alpha_k}{2(k+1)-d}
	\Lambda^{d-2k-2}(-\nabla^2)^k \\ -\alpha_{\frac{d}{2}-1}
	(-\nabla^2)^{\frac{d}{2}-1}
	\log\left(\frac{-\nabla^2}{\Lambda^2}\right)\,.  \label{kernel_even_d}
	\end{multline} Thus, all the nonlocal structure is logarithmic:
	\begin{equation}
		\int_{0}^{\infty}\frac{\mathrm{d}s}{s^{d/2}}\,\alpha(-s\nabla^2)
	= -\alpha_{\frac{d}{2}-1} (-\nabla^2)^{\frac{d}{2}-1}
\log\left(\frac{-\nabla^2}{\mu^2}\right) +\text{local divergent terms}\,,
\end{equation} where we have added and subtracted the local term
$\alpha_{\frac{d}{2}-1}\log(\mu^2)(-\nabla^2)^{\frac{d}{2}-1}$, with $\mu$ a
finite energy scale. Therefore, the renormalised nonlocal effective action reads
\begin{equation}
    \Gamma_{\mathrm{NL}}^{(2)} = \frac{1}{2}\kappa_d
    \int_{\partial\mathcal{M}}  K(-\nabla^2)^{\frac{d}{2}-1}
    \log\left(\frac{-\nabla^2}{\mu^2}\right)K\,,
    \label{eff_action_even}
    \end{equation}
where we have defined $\kappa_d:=\alpha_{\frac{d}{2}-1}/(4\pi)^{d/2}$ for even $d$ to
streamline the notation.

\subsection{Odd-dimensional boundary}

For odd $d$, we obtain \begin{equation}
	\int_{\frac{1}{\Lambda^2}}^{\infty}\frac{\mathrm{d}s}{s^{d/2}}\,\alpha(-s\nabla^2)=\text{const.}\times(-\nabla^2)^{\frac{d}{2}-1}-\sum_{k=0}^{\frac{d-3}{2}}\frac{2\,\alpha_k}{2(k+1)-d}\Lambda^{d-2k-2}(-\nabla^2)^k\,.
\label{kernel_odd_d} \end{equation}
Again, the divergent terms are proportional
to integer powers of $\nabla^2$, and therefore give local contributions to the
unrenormalised effective action. However, in this case the nonlocal structure is
encoded in the fractional power of the Laplace-Beltrami operator appearing in
the first term on the right-hand side of Eq.~\eqref{kernel_odd_d}. Its
coefficient cannot be determined from the ultraviolet expansion alone, because
it depends on the upper limit of the proper-time integral. To compute this
constant, one needs to know the function $\alpha(\xi)$ for all values of its
argument; the first terms of its small-$\xi$ expansion, unlike in the
even-dimensional case, are not sufficient. Thus, in this case we can write the
renormalised nonlocal effective action as
\begin{equation}
\Gamma_{\mathrm{NL}}^{(2)} = \frac{1}{2}\kappa_d
\int_{\partial\mathcal{M}} K(-\nabla^2)^{\frac{d}{2}-1}K\,,
\label{eff_action_odd}
\end{equation}
where the constant $\kappa_d$ for odd $d$ is given by
\begin{equation}
\label{kappa_odd_d} \kappa_d =-\frac{1}{(4\pi)^{d/2}}
\int_{0}^{\infty}\frac{\mathrm{d}\xi}{\xi^{d/2}}\left(\alpha(\xi)-\sum_{k=0}^{\frac{d-3}{2}}\alpha_k\,\xi^k\right)\,.
\end{equation}

In both even and odd dimensions, we obtained the nonlocal contributions to the
effective action by introducing a UV cutoff in the unrenormalised action. In
other words, in either case the form of the nonlocal operator is encoded in the
divergent contributions, although the proportionality constant $\kappa_d$ in
Eq.~\eqref{eff_action_odd} cannot be determined solely from the divergent terms
in the heat-kernel expansion. Therefore, the expressions in
Eqs.~\eqref{eff_action_even} and \eqref{eff_action_odd} encode the full nonlocal
structure of the effective action at quadratic order in the extrinsic curvature,
even though they were obtained from Eq.~\eqref{eff_action_2_1}, which describes
only the UV sector of the complete effective action.

With the coefficients in Eqs.~\eqref{alphaD} and \eqref{alphaN}, we can express
the nonlocal part of the effective action for both Dirichlet and Neumann
boundary conditions on boundaries of dimensions $d=2$ and $d=4$. Higher
even-dimensional boundaries require additional heat-kernel coefficients, whose
computation becomes increasingly involved. On the other hand, for
odd-dimensional boundaries, the heat-kernel coefficients alone are not enough to
determine the overall coefficient of the nonlocal effective action. However, in
the next section we will show that the coefficients in arbitrary dimension can
be determined by comparison with previous results \cite{Bruno:Dirichlet,
Bruno:Neumann}, where the nonlocal part of the effective action was computed for
a more restricted class of boundaries.

%====================================================================
\section{Monge patch}\label{sec:monge}

In this section we unify the results of this work with those of
Refs.~\cite{Bruno:Dirichlet,Bruno:Neumann}, showing that the coefficients in
Eqs.~\eqref{eff_action_even} and \eqref{eff_action_odd} can be obtained from the
nonlocal effective action of a massless scalar field in $d+1$ dimensions subject
to Dirichlet or Neumann boundary conditions on a hypersurface parametrised as a
Monge patch, with the field defined on both sides of the hypersurface. Comparing
that setup to the present one, in which a massless scalar field on flat
spacetime is restricted to one side of the same hypersurface, the only
difference is the doubling of the available bulk volume, i.e. above and below the surface,  which produces an
overall factor of $\tfrac{1}{2}$ between the two effective actions, as we now
show. 

To compare the coefficients of the two effective actions for even $d$, we write
\eqref{eff_action_even} in the case that the boundary admits
a single global Monge patch. Following the conventions of
Refs.~\cite{Bruno:Dirichlet,Bruno:Neumann}, we use the first $d$ spacetime
components of $x^\mu$ as parameters, $y=(x_1,\cdots,x_{d})$, and define the
Monge patch by \begin{equation}\label{eq:defmonge}
\Sigma\;:\; x_{d+1} \,=\, \psi(y) \;,
\end{equation}
where $\psi(y)$ specifies the ``height'' of the surface, measured with respect to the hyperplane $x_{d+1} = 0$, and the $y$ parameters are left as Cartesian coordinates. Since the target effective actions are of second order in the height function $\psi$, we adapt Eq.~\eqref{eff_action_even} to the same regime. The scalar extrinsic curvature for a Monge patch is

\begin{equation} \label{def:extrinsic-curvature-monge}
K=\partial_a\left(\frac{\partial^a \psi }{\sqrt{1+\partial_b \psi \, \partial^b
\psi\,}}\right)\, \end{equation}
where indices are contracted with the Euclidean metric $\delta_{ab}$. We can expand Eq.~\eqref{def:extrinsic-curvature-monge} in powers of $\psi$ as

\begin{equation} K= \partial^2 \, \psi - \frac{1}{2} \partial^2\, \psi
\,\partial_a \psi \partial^a \psi - \partial^a \psi \partial^b \psi
\partial_{a}\partial_{b} \psi+ \cdots \end{equation}
where the leading term is first order in $\psi$, namely $\partial^2\psi$, with $\partial^2$ the Euclidean $d$-dimensional Laplacian in Cartesian coordinates. 

Since Eq.~\eqref{eff_action_even} is of order $2$ in $K$, and the leading term in
$K$ is of order $1$ in $\psi$, it suffices to retain $K \approx \partial^2\psi$
inside that expression to obtain the order-$\psi^2$ part of the effective
action; equivalently, we identify $\nabla^2 \to \partial^2$. Using integration
by parts, we then write $K\,(-\partial^2)^{\frac{d}{2}-1} \log(-\partial^2) K$
as $\psi\, (-\partial^2)^{\frac{d}{2}+1} \log(-\partial^2) \psi$, valid inside
the integral. Therefore, for the Monge patch in Eq.~\eqref{eq:defmonge} and up to second
order in $\psi$, the effective action with even $d$ in Eq.~\eqref{eff_action_even} reads

\begin{equation}\label{eff_action_even_Monge} \Gamma_{\mathrm{NL}}^{(2)} =
\frac{1}{2}\kappa_d^{(\mathrm{D},\mathrm{N})} \int d^dy\, \left[
\psi(y)\;(-\partial^2)^{\frac{d}{2}+1}
\log\left(\frac{-\partial^2}{\mu^2}\right)\psi(y) \right]\,, \end{equation}
where the integral is over the full Cartesian range of the parameters $y$. 

The target formulas are, for even values of $d$,

\begin{equation}\label{eff_action_even_FG}
    \Gamma^{(2)}_{[FG]}= \frac{1}{2} \eta^{(\mathrm{D},\mathrm{N})}_d \int \frac{d^d k}{(2\pi)^d} \, |\tilde{\psi}(k)|^2 (k^2)^{\frac{d}{2}+1} \log\left(\frac{k^2}{\mu^2}\right)\,, 
\end{equation}
where we identify the correct functional structure using the Fourier transform convention\linebreak $\tilde{\psi}(k)=\int \mathrm{d}^d y\,e^{-i k \cdot y}\psi(y)$. Having
\begin{align} 
     \eta_d^{(\mathrm{D})}&=\frac{(-1)^{\frac{d}{2}+1}}{(4\pi)^{\frac{d}{2}+1}} \frac{(\Gamma(\frac{d+1}{2}))^2}{(\frac{d}{2}+1)! \,d\,!} \\ 
     \eta_d^{(\mathrm{N})}&= \eta_d^{(\mathrm{D})} +\frac{(-1)^{(\frac{d}{2}+1)}}{(d/2)!} (d-1) \frac{(\Gamma(\frac{d+1}{2}))^2}{2^{d+1} \pi^{\frac{d+2}{2}}\Gamma(d+1)}\,,
\end{align}
we verified, for $d=2$ and $d=4$, that
\begin{equation}
\label{coeff-relation-even}
    \kappa^{(\mathrm{D},\mathrm{N})}_{d}=\frac{1}{2}\eta^{(\mathrm{D},\mathrm{N})}_{d}
\end{equation}
is satisfied, and the factor $\tfrac{1}{2}$ is due to the difference between defining the theory on both sides of the boundary and on a single side. Moreover, the relation in Eq.~\eqref{coeff-relation-even} holds for all even values of $d$, since the manipulation that takes Eq.~\eqref{eff_action_even_FG} into Eq.~\eqref{eff_action_even_Monge} does not affect the prefactor of the integral, regardless of the value of $d$.

For an odd-dimensional boundary, we have

\begin{equation} \Gamma^{(2)}_{[FG]}= \frac{1}{2} \eta^{(\mathrm{D},\mathrm{N})}_d \int \frac{d^d
k}{(2\pi)^d} \, |\tilde{\psi}(k)|^2 (k^2)^{\frac{d}{2}+1} \,, \end{equation} 
with 

\begin{align}
    \eta_d^{(\mathrm{D})}&=\frac{(-1)^{\frac{d+1}{2}}}{(2\pi)^{\frac{d+1}{2}}}
	\frac{((\frac{d-1}{2})!)^2}{d\,! (d+2)!!} \\ \eta_d^{(\mathrm{N})}&= \eta_d^{(\mathrm{D})}
	+ (d-1) \frac{(\Gamma(\frac{d+1}{2}))^2 \Gamma(-\frac{d}{2})}{2^{d+1}
	\pi^{\frac{d+2}{2}}\Gamma(d+1)}\,.
\end{align}

Although $\kappa_d$ in Eq.~\eqref{kappa_odd_d} cannot be evaluated directly without the explicit form of $\alpha(\xi)$, its value can be fixed by using the Monge-patch case in the approximation $K \approx \partial^2 \psi$. This leads to the same coefficient relation as in Eq.~\eqref{coeff-relation-even}. Therefore, Eq.~\eqref{coeff-relation-even} holds for both even and odd boundary dimensions.

%====================================================================
\section{Examples}\label{examples}

In this section we consider an oscillating deformed ring in $2+1$ dimensions and
an oscillating deformed sphere in $3+1$ dimensions. The two cases differ in an
essential way. In Euclidean formulation, the unperturbed (static) ring
corresponds to a cylinder, which has vanishing intrinsic curvature, and the
structures $K^2$ and $K_{ab}K^{ab}$ are trivially proportional at the background
level. The sphere, on the other hand, has nontrivial intrinsic curvature already
at the unperturbed level, so that the reduction of $K_{ab}K^{ab}$ to $K^2$ in
Eq.~\eqref{eq:KabKab-to-K2} relies essentially on the presence of the
nonlocal operator inside the integral.

\subsection{Oscillating ring}\label{sec:ring}
Let us compute the nonlocal effective action for a cylindrical boundary in the
case $d=2$ for both Dirichlet and Neumann boundary conditions. Using cylindrical
coordinates $\{r,\theta,z\}$ in the bulk $\mathbb{R}^3$, we consider the
boundary surface $\Sigma$ defined by
\begin{equation}
    \Sigma\;:\;\; r=R\,(1+\epsilon\,\cos(n\theta)\cos(mz))\,,
\end{equation}
where $\epsilon\ll1$ is a small dimensionless perturbative parameter, and $R$ is
the radius of the unperturbed cylinder. Choosing $\{\theta,z\}$, with
$0\leq\theta<2\pi$ and $z\in\mathbb{R}$ as coordinates on $\Sigma$, the scalar extrinsic curvature is
\begin{equation}
  K=K^{(0)}+\epsilon\, K^{(1)}+\mathcal{O}(\epsilon^2)\,,
  \label{cyl00}
\end{equation}
where the superscript indicates the order in the expansion in $\epsilon$, 
\begin{equation}
    K^{(0)}=-\frac{1}{R}\,,
    \qquad
    K^{(1)}=\frac{1}{R}
    \left(1-n^2-m^2R^2\right)
    \cos(mz)\cos(n\theta)\,,
\end{equation}
and we have adopted here the convention that the normal points inwards. 

The contribution to the nonlocal effective action from Eq.~\eqref{eff_action_even} reads
\begin{equation} \label{def_gamma_cylinder}
  \Gamma_{\mathrm{NL}}^{(2)}=\frac{\kappa_2^{(\mathrm{D},\mathrm{N})}}{2}\int_{0}^{2\pi}\mathrm{d}\theta\int_{-\infty}^{\infty}\mathrm{d}z\,\sqrt{h}\,K\log\left(\frac{-\nabla^2}{\mu^2}\right)K\,,
\end{equation}
where we expand in powers of $\epsilon$ by adding a second superscript denoting the order in $\epsilon$:
\begin{equation}
\Gamma^{(2)}_{\mathrm{NL}}=\Gamma^{(2,0)}_{\mathrm{NL}}+\Gamma^{(2,1)}_{\mathrm{NL}}+\Gamma^{(2,2)}_{\mathrm{NL}}+\cdots\,.
  \label{cyl_gammap}
\end{equation}
Using the convention $\log\left(-\nabla^2/\mu^2\right)\,\text{constant}=0$, together with the self-adjointness of $\nabla^2$ with respect to the usual inner product on $\Sigma$, so
\begin{equation}
  \int_{0}^{2\pi}\mathrm{d}\theta\int_{-\infty}^{\infty}\mathrm{d}z\,\sqrt{h}\,\varphi\log\left(\frac{-\nabla^2}{\mu^2}\right)\psi
  =
  \int_{0}^{2\pi}\mathrm{d}\theta\int_{-\infty}^{\infty}\mathrm{d}z\,\sqrt{h}\,\psi\log\left(\frac{-\nabla^2}{\mu^2}\right)\varphi\,,
  \label{cyl_selfadj}
\end{equation}
for any test functions $\varphi$ and $\psi$, we find that the first nontrivial contribution to Eq.~\eqref{def_gamma_cylinder} comes from the term quadratic in $K^{(1)}$, namely
 \begin{equation}
       \Gamma_{\mathrm{NL}}^{(2,2)}=\frac{\kappa_2^{(\mathrm{D},\mathrm{N})}}{2}\int_{0}^{2\pi}\mathrm{d}\theta\int_{-\infty}^{\infty}\mathrm{d}z\,\sqrt{h^{(0)}}\,K^{(1)}\log\left(\frac{-{\nabla^2}^{(0)}}{\mu^2}\right)K^{(1)}\,,
\end{equation}
where $\sqrt{h^{(0)}}=R$ and ${\nabla^2}^{(0)}=R^{-2}\partial^{2}_{\theta}+\partial^{2}_{z}$ are used to remain at fixed order in $\epsilon$.

Observing that $K^{(1)}$ is an eigenfunction of $-{\nabla^2}^{(0)}$ with eigenvalue $\frac{n^2}{R^2}+m^2$, we have
\begin{equation}
  \Gamma_{\mathrm{NL}}^{(2,2)}=\frac{\kappa_2^{(\mathrm{D},\mathrm{N})}}{2}\frac{\epsilon^2\pi L\left(n^2+m^2R^2-1\right)^2}{2R}\,\log\left(\frac{\frac{n^2}{R^2}+m^2}{\mu^2}\right)\,,
  \label{cyl_eff2E}
\end{equation}
where $L=\int_{-\infty}^{\infty} dz$ is a total length factor. To pass to Lorentzian signature, we perform the Wick rotation $-\Gamma_{\mathrm{NL}}^{(2,2)} \to i  \Gamma_{\mathrm{NL}}^{(2,2)}$, by setting $z=it$, with $t$ the Lorentzian time, and performing the replacement $m^2 \to -\omega^2-i0^+$ with the corresponding Feynman prescription. Therefore,
\begin{equation} \label{cyl_eff2L}
  \Gamma^{(2,2)}_{\mathrm{NL}}=-\frac{\kappa_2^{(\mathrm{D},\mathrm{N})}}{2}\frac{\epsilon^2\pi T\left(n^2-\omega^2R^2-1\right)^2}{2R}\,\log\left(\frac{\frac{n^2}{R^2}-\omega^2-i0^+}{\mu^2}\right)\,,
\end{equation}
where denotes the total time interval. The effective action in Eq.~\eqref{cyl_eff2L} corresponds to the quantum field theory of a massless scalar field in flat $(2+1)$-dimensional spacetime, living in the inner or outer region of a ring of radius $R$ with $n$ sinusoidal deformations oscillating at frequency $\omega$, with Dirichlet or Neumann boundary conditions imposed at the boundary.

The range of validity of the approximation is given by $\epsilon\ll1$ and $\nabla\nabla K\gg K^3$, which in terms of the physical parameters become $ n\gg1$ or $\omega R\gg1$. Within this range, taking the imaginary part of Eq.~\eqref{cyl_eff2L} we get
\begin{equation}
 \frac{\mathrm{Im}\left[\Gamma^{(2,2)}_{\mathrm{N L}}\right]}{T}=\frac{\kappa_2^{(\mathrm{D},\mathrm{N})}}{2}\frac{\epsilon^2\pi^2 \left(n^2-\omega^2R^2-1\right)^2}{2R}\,\Theta\left(\omega R-n\right)\,,
  \label{imGamma}
\end{equation}
where $\kappa_2^{(\mathrm{D})}=\frac{1}{512\pi}$ and $\kappa_2^{(\mathrm{N})}=\frac{5}{512\pi}$.

\subsection{Oscillating multipolar sphere}\label{sec:multipole}

We turn now to a closed boundary which cannot be described by a single Monge
patch and on which the background extrinsic geometry is itself nontrivial:
a $2$-sphere of radius $R$ in $3{+}1$ dimensions, undergoing a small
time-dependent deformation along a single spherical-harmonic mode,
\begin{equation}
  \Sigma_t\;:\;\; r=\rho(t,\Omega)=R\bigl[1+\epsilon\,f(t)\,Y_{\ell m}(\Omega)\bigr]\,,
  \qquad \epsilon\ll 1\,,
  \label{spherical_surface}
\end{equation}
with $\Omega=(\theta,\phi)$ and $Y_{\ell m}$ a real spherical harmonic, taken
as an eigenfunction of the unit-sphere Laplacian,
$\nabla_{\Omega}^{2}Y_{\ell m}=-\ell(\ell+1)Y_{\ell m}$. The Euclidean
worldtube is $\partial\mathcal{M}=S^{2}_{R}\times\mathbb{R}_{\tau}$, so that
$d=3$ and we are in the odd-dimensional case described by
Eq.~\eqref{eff_action_odd}. The case $\ell=0$ corresponds to the breathing
mode considered, by a different method, in Ref.~\cite{sphere} and reproduced
in the appendix; here we focus on the genuinely nontrivial multipolar modes
$\ell\geq 1$.

A direct computation, performed using the level-set formula for the mean
curvature of the worldtube embedded in flat Euclidean $\mathbb{R}^{4}$ (or, in
Lorentzian signature, in $\mathbb{R}^{1,3}$), yields, to first order in
$\epsilon$,
\begin{equation}\label{eq:K-sphere-general}
  K\;=\;\frac{2}{R}\;-\;\frac{\epsilon}{R}
  \Bigl[\,R^{2}\,\partial_{t}^{2}h \;+\; 2\,h \;+\; \nabla_{\Omega}^{2}h\,\Bigr]
  \;+\;\mathcal{O}(\epsilon^{2})\,,
\end{equation}
where $h(t,\Omega):=f(t)\,Y_{\ell m}(\Omega)$ is the (rescaled) deformation
profile.\footnote{We are using the sign convention in
which a static sphere bounding the bulk from outside has $K^{(0)}=+2/R$;
the action depends bilinearly on $K$, so the overall sign is irrelevant.}
Specialising to a harmonic mode $h=f(t)Y_{\ell m}$,
\begin{equation}\label{eq:K1-multipole}
  K^{(1)}(t,\Omega)\;=\;-\frac{1}{R}\Bigl[R^{2}\ddot f(t)
  -(\ell-1)(\ell+2)\,f(t)\Bigr] Y_{\ell m}(\Omega)\,,
\end{equation}
where we have used the algebraic identity
$\ell(\ell+1)-2=(\ell-1)(\ell+2)$, which exhibits the well-known fact that
the dipole modes ($\ell=1$) carry no static contribution to $K^{(1)}$, since a
rigid translation of the sphere leaves its extrinsic geometry unchanged,
and that the lowest genuinely shape-changing oscillation is the quadrupole,
$\ell=2$, where $(\ell-1)(\ell+2)=4$.

To find the nonlocal effective action, for $d=3$, the relevant building block 
is Eq.~\eqref{eff_action_odd}, considering its first non-trivial contribution in orders of $\epsilon$, namely
\begin{equation}
  \Gamma_{\mathrm{NL}}^{(2,2)}
  \;=\;\frac{\kappa_{3}^{(\mathrm{D},\mathrm{N})}}{2}\,
  \int_{\partial\mathcal{M}}\!\sqrt{h^{(0)}}\;K^{(1)}\left({-\nabla^{2}}^{(0)}\right)^{1/2}K^{(1)}\,,
\end{equation}
with the coefficients
\begin{equation}\label{eq:kappa3-values}
  \kappa_{3}^{(\mathrm{D})}=\frac{1}{720\pi^2}\,,
  \qquad
  \kappa_{3}^{(\mathrm{N})}=\frac{11}{720\pi^2}\,,
\end{equation}
read off from Eq.~\eqref{coeff-relation-even} (extended to odd $d$), and again we're using $\sqrt{h^{(0)}}$ and ${\nabla^2}^{(0)}$ to fix the order in $\epsilon$. The
Laplace-Beltrami operator on the worldtube
$\partial\mathcal{M}=S^{2}_{R}\times\mathbb{R}_{t}$ is, in Lorentzian
signature,
\begin{equation}
  \Box\;=\;\partial_{t}^{2}-\frac{1}{R^{2}}\,\nabla_{\Omega}^{2}\,,
\end{equation}
and its eigenvalues on plane waves $e^{-i\omega t}Y_{\ell m}(\Omega)$ are
$\lambda_{\ell}(\omega)=-\omega^{2}+\ell(\ell+1)/R^{2}$. With the Feynman
$-i\,0^{+}$ prescription,
\begin{equation}\label{eq:branch-cut}
  \Box^{1/2}\;e^{-i\omega t}Y_{\ell m}\;=\;
  \begin{cases}
     +\sqrt{\,\omega_{\ell}^{2}-\omega^{2}\,}\;e^{-i\omega t}Y_{\ell m}\,,
        & \omega^{2}<\omega_{\ell}^{2}\,, \\[2pt]
     -i\sqrt{\,\omega^{2}-\omega_{\ell}^{2}\,}\;e^{-i\omega t}Y_{\ell m}\,,
        & \omega^{2}>\omega_{\ell}^{2}\,,
  \end{cases}
\end{equation}
with the multipole-dependent gap
\begin{equation}\label{eq:gap}
  \omega_{\ell}\;:=\;\frac{\sqrt{\ell(\ell+1)}}{R}\,.
\end{equation}
The branch cut at $\omega=\omega_\ell$ is the geometric origin of the imaginary
part of the effective action and, hence, of the radiation rate: the multipole
$\ell$ radiates only when the oscillation frequency exceeds $\omega_\ell$.

%\paragraph{Imaginary part of the effective action.}
Inserting Eq.~\eqref{eq:K1-multipole} into the effective action and using
Eq.~\eqref{eq:branch-cut} together with the orthogonality of the spherical
harmonics, $\int_{S^{2}}|Y_{\ell m}|^{2}\,d\Omega=1$, and
$\sqrt{h^{(0)}}\,d^{3}y=R^{2}\sin\theta\,d\theta\,d\phi\,dt$, one finds, for a
monochromatic deformation $f(t)=\cos(\omega t)$ over a window of duration
$T\gg 1/\omega$,
\begin{equation}\label{eq:ImG-multipole}
  \mathrm{Im}\,\left[\Gamma_{\mathrm{NL}}^{(2,2)}\right]
  \;=\;\frac{\kappa_{3}^{(\mathrm{D},\mathrm{N})}}{4}\;
  \epsilon^{2}\,T\,\bigl[\Lambda_{\ell}(\omega)\bigr]^{2}\,
  \sqrt{\,\omega^{2}-\omega_{\ell}^{2}\,}\;
  \Theta(\omega-\omega_{\ell})\,,
\end{equation}
with
\begin{equation}\label{eq:Lambda-multipole}
  \Lambda_{\ell}(\omega)\;:=\;R^{2}\omega^{2}\,-\,(\ell-1)(\ell+2)\,.
\end{equation}
Equivalently, using
$\kappa_{3}^{(\mathrm{D})}/4=1/(2880\pi^{2})$ and
$\kappa_{3}^{(\mathrm{N})}/4=11/(2880\pi^{2})$,
\begin{equation}
  \mathrm{Im}\,\left[\Gamma_{\mathrm{NL}}^{(2,2)}\right]
  \;=\;\frac{c_{(\mathrm{D},\mathrm{N})}}{\pi^{2}}\,\epsilon^{2}\,T\,
  \bigl[R^{2}\omega^{2}-(\ell-1)(\ell+2)\bigr]^{2}\,
  \sqrt{\,\omega^{2}-\omega_{\ell}^{2}\,}\,,\quad
  \omega>\omega_{\ell}\,,
\end{equation}
with $c_{\mathrm{D}}=1/2880$ and $c_{\mathrm{N}}=11/2880$.

%\paragraph{Mean number of created particles.}
The mean number of created particles can be computed using the relation $\mathrm{Im}\,\Gamma\approx\langle N\rangle/4$ (see the \nameref{app:bogoliubov}),
\begin{equation}
  \frac{\langle N\rangle}{T}\;\approx\;\frac{4 c_{(\mathrm{D},\mathrm{N})}}{\pi^{2}}\,
  \epsilon^{2}\,\bigl[R^{2}\omega^{2}-(\ell-1)(\ell+2)\bigr]^{2}\,
  \sqrt{\,\omega^{2}-\omega_{\ell}^{2}\,}\,,\qquad \omega>\omega_{\ell}\,.
  \label{sphere_mean_number}
\end{equation}

\begin{figure}[t]
  \centering
  \includegraphics[width=0.78\linewidth]{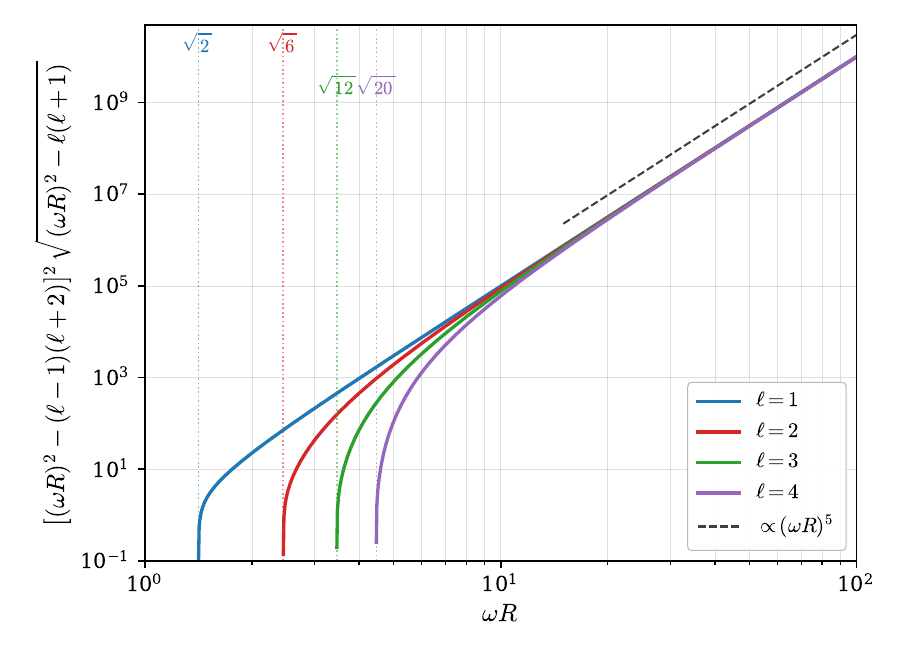}
  \caption{Dimensionless mean number of created particles per unit time,
    $\bigl[(\omega R)^{2}-(\ell-1)(\ell+2)\bigr]^{2}\sqrt{(\omega R)^{2}-\ell(\ell+1)}$,
    as a function of $\omega R$, for the multipoles $\ell=1,2,3,4$.
    The full rate is recovered by multiplication by
    $4 c_{(\mathrm{D},\mathrm{N})}\,\epsilon^{2}/(\pi^{2}R)$ with
    $c_{\mathrm{D}}=1/2880$ and $c_{\mathrm{N}}=11/2880$; the Neumann result
    is therefore obtained from the Dirichlet curve by an overall factor of
    $11$, independent of $\ell$. Coloured dotted vertical lines mark the
    multipole-dependent gap frequencies $\omega_{\ell}R=\sqrt{\ell(\ell+1)}$,
    below which the corresponding mode does not radiate. The dashed grey line
    indicates the universal high-frequency asymptote $\propto(\omega R)^{5}$
    onto which all multipoles converge for $\omega R\gg\sqrt{\ell(\ell+1)}$,
    recovering the scaling expected for a two-dimensional oscillating mirror
    in $3{+}1$ dimensions.}
  \label{fig:multipole_rate}
\end{figure}

Several features of this example, summarised in Fig.~\ref{fig:multipole_rate}, deserve comment.
\begin{enumerate}
  \item The mode $\ell$ radiates only
        for $\omega>\omega_{\ell}=\sqrt{\ell(\ell+1)}/R$. The threshold is
        the angular eigenvalue of the worldtube Laplacian; geometrically,
        below it the boundary perturbation cannot supply enough on-shell
        momentum to the pair of emitted quanta. Note that
        $\omega_{0}=0$ (no threshold for the breathing mode),
        $\omega_{1}=\sqrt{2}/R$, $\omega_{2}=\sqrt{6}/R$, etc.
  \item For $\ell=1$,
        Eq.~\eqref{eq:K1-multipole} reduces to
        $K^{(1)}=-R\,\ddot f(t)\,Y_{1m}(\Omega)$. The static piece vanishes
        identically because a rigid sphere translation preserves the
        extrinsic geometry, and only the kinematic $\ddot f$ piece contributes.
        Consistently, $\Lambda_{1}(\omega)=R^{2}\omega^{2}$ vanishes at
        $\omega=0$.
  \item The lowest genuine shape oscillation is $\ell=2$,
        with $\omega_{2}=\sqrt{6}/R$ and
        $\Lambda_{2}(\omega)=R^{2}\omega^{2}-4$. For Dirichlet and Neumann
        conditions,
        \begin{align}
          \left.\frac{\langle N\rangle}{T}\right|_{\ell=2}^{(\mathrm{D})}
          &=\frac{1}{720\pi^{2}}\,\epsilon^{2}\,
            \bigl(R^{2}\omega^{2}-4\bigr)^{2}\,
            \sqrt{\,\omega^{2}-6/R^{2}\,}\,, \\[2pt]
          \left.\frac{\langle N\rangle}{T}\right|_{\ell=2}^{(\mathrm{N})}
          &=\frac{11}{720\pi^{2}}\,\epsilon^{2}\,
            \bigl(R^{2}\omega^{2}-4\bigr)^{2}\,
            \sqrt{\,\omega^{2}-6/R^{2}\,}\,.
        \end{align}
        The Neumann rate is $11$ times the Dirichlet rate, the same universal
        ratio dictated by $\kappa_{3}^{(\mathrm{N})}/\kappa_{3}^{(\mathrm{D})}$
        and therefore independent of $\ell$.

  \item For $\omega R\gg 1$, all multipoles
        reach the regime \mbox{$\langle N\rangle/T\simeq (4 
        c_{(\mathrm{D},\mathrm{N})}/\pi^{2})\,
         \epsilon^{2}R^{4}\omega^{5}$}, recovering the universal
        $\omega^{5}$ scaling expected for emission from a $2$-dimensional
        oscillating mirror in $3{+}1$ dimensions.
\end{enumerate}

The result above complements the breathing-sphere computation of the appendix:
in the bulk-mode language of \cite{sphere}, only $\ell=0$ contributes to the
$\delta(W_{\ell n k}-\omega)$ at leading order in $\epsilon$, in agreement with
the threshold $\omega_{0}=0$ of our formula.

%====================================================================
\section{Conclusions}\label{sec:conclusions}

We have constructed the nonlocal effective action of a free massless scalar
field in flat $(d{+}1)$-dimensional space-time, in the presence of either
Dirichlet or Neumann boundary conditions on a smooth, codimension-one
hypersurface $\partial\mathcal{M}$, working at quadratic order in the
extrinsic curvature tensor $K_{ab}$. Two features distinguish this construction
from previous treatments of the same physical setting. First, the result is
manifestly covariant under diffeomorphisms of $\partial\mathcal{M}$: it is
expressed entirely in terms of the geometric objects $K_{ab}$, the induced
metric $h_{ab}$, the Laplace-Beltrami operator $\nabla^{2}$ on the
boundary, and no embedding-specific or coordinate-dependent quantity (such as
a height function $\psi$) appears. Second, the construction does not require
a global Monge-patch parametrisation, and therefore extends to surfaces with
overhang or non-trivial topology that lie outside the regime of
Refs.~\cite{Bruno:Dirichlet,Bruno:Neumann}.

The structure of the result reflects the parity of the boundary manifold 
dimension. In
even $d$, the kernel is $(-\nabla^{2})^{d/2-1}\log(-\nabla^{2}/\mu^{2})$, with
an overall coefficient $\kappa_{d}$ that is fully determined by the
half-integer-indexed boundary heat-kernel coefficient $b_{(d-1)/2+1}$. In odd
$d$, the kernel is the fractional power $(-\nabla^{2})^{d/2-1}$, and its
overall coefficient is not accessible from the divergent terms of the
heat-kernel expansion alone: it requires either a full knowledge of the
function $\alpha(\xi)$ defined in Eq.~\eqref{eff_action_2_2}, or, as we showed
in Section~\ref{sec:monge}, a comparison with the Monge-patch results of
Refs.~\cite{Bruno:Dirichlet,Bruno:Neumann}. The latter route fixes
$\kappa_{d}$ in arbitrary $d$ and provides a non-trivial bridge between
the geometric and the height-function formulations: the Monge-patch result is
recovered as a special case of our covariant expression, with the overall
factor of $\tfrac{1}{2}$ accounting for whether the field lives on one or
both sides of the boundary. The reduction $\int K_{ab}K^{ab}=\int K^{2}$ at
order $K^{2}$, derived through the Codazzi identity and a definite
prescription for the inverse Laplacian, plays a central role in the unification.

Two explicit applications illustrate the formalism. The oscillating ring in
$2{+}1$ dimensions is a case in which the relevant  dynamics is encoded in the 
perturbation around a flat induced metric. The oscillating sphere in
$3{+}1$ dimensions, by contrast, has nonvanishing intrinsic curvature already
at the unperturbed level, and provides a test of the covariant reduction
of $K^{ab}K_{ab}$ to $K^{2}$ when the background is intrinsically curved.
For a multipolar deformation $r=R[1+\epsilon\cos(\omega t)
Y_{\ell m}(\Omega)]$, the imaginary part of the effective action displays a
mode-by-mode threshold $\omega_{\ell}=\sqrt{\ell(\ell+1)}/R$ inherited from
the angular spectrum of the worldtube, and a universal Neumann-to-Dirichlet
ratio of the radiation rate equal to $\kappa_{3}^{(\mathrm{N})}/
\kappa_{3}^{(\mathrm{D})}=11$, independent of the multipole (the same ratio appears  when computing the dynamical Casimir effect for TM and TE modes of the electromagnetic field in a plane cavity \cite{Mundarain}). The dipole sector
$\ell=1$ is silent at static order in the deformation, consistent with the
geometric fact that a rigid translation of the sphere leaves $K_{ab}$
unchanged.

Several natural extensions might be within the reach of the same techniques.
Pushing the geometric expansion to cubic order in $K_{ab}$ is the immediate
next step: the integer-indexed coefficient $b_{2}$ is fully tabulated, but
introduces independent invariants: $K^{3}$, $KK^{ab}K_{ab}$, and
$K_{a}{}^{b}K_{b}{}^{c}K_{c}{}^{a}$: that mix nontrivially under integration
by parts when the boundary has nonzero intrinsic curvature, and require a
careful generalisation of the present reduction. Beyond cubic order, the
heat-kernel route alone is insufficient and the analysis must be supplemented
by a direct calculation of $\alpha(\xi)$, accessible in principle from
explicit propagator computations in geometries of high symmetry. A second
direction is the inclusion of richer field content like fermions and gauge
fields, each with its own set of boundary heat-kernel coefficients, required 
for the  dynamical Casimir effect of the electromagnetic
field. A third is the combination of the present geometric setup with a
curved bulk, relevant for accelerated mirrors in cosmological backgrounds and
for boundary contributions to semiclassical gravity. We hope to address some of 
these directions in future work.

%====================================================================
\section*{Acknowledgments}

This research was supported by Consejo Nacional de Investigaciones Científicas y Técnicas (CONICET).

%====================================================================
\appendix

\section*{Appendix}\label{app:bogoliubov}
\subsection*{Relation between \texorpdfstring{$\mathrm{Im}\,\Gamma$}{Im Gamma} and Bogoliubov coefficients}

As an explicit test of our result in Eq.~\eqref{sphere_mean_number} we compute  the imaginary part of the effective action for an oscillating sphere, using the results of Ref.~\cite{sphere}, where the authors obtained the mean number of created particles through Bogoliubov transformations.

First, we relate the mean number of created particles to the imaginary part of the effective action. Let $\alpha$ and $\beta$ be the matrices of Bogoliubov coefficients relating the \textit{in} and \textit{out} Fock spaces, whose vacuum states are $|0_{\text{in}}\rangle$ and $|0_{\text{out}}\rangle$, respectively. If $N_i$ is the number operator associated with the $i$-th mode of the \textit{out} Fock space, then the total number of particles evaluated in the \textit{in} vacuum is \cite{birrell1982}
\begin{equation}
    \langle N\rangle
    \equiv
    \sum_i\langle 0_{\text{in}}|N_i|0_{\text{in}}\rangle
    =
    \mathrm{Tr}(\beta\beta^{\dagger})\,.
\end{equation}
To relate this quantity to the imaginary part of the Lorentzian effective action $\Gamma$, we express the latter in terms of the Bogoliubov coefficients through the relation \cite{DEWITT1975}
\begin{equation}
    e^{i\Gamma}=(\det\alpha)^{-1/2}\,.
    \label{ape1}
\end{equation}
Thus, Eq.~\eqref{ape1} gives
\begin{equation}
    \mathrm{Im}\,\Gamma
    =
    \frac{1}{4}\mathrm{Tr}\left[\log\left(1+\beta\beta^{\dagger}\right)\right]
    \quad\Longrightarrow\quad
    \mathrm{Im}\,\Gamma\approx\frac{1}{4}\langle N\rangle\,,
\end{equation}
where the approximation holds when the number of produced particles is perturbatively small.

Ref.~\cite{sphere} considers a massless scalar QFT in the interior of a sphere with radius $a(t)$, with Dirichlet boundary conditions imposed at the surface. The time-dependent radius is assumed to be
\begin{equation}
    a(t)=
    \begin{cases}
        R\left[1+\epsilon\,Y_{00}(\Omega)\sin(\omega t)\right], & 0\leq t\leq T\,, \\
        R, & t<0,\quad t>T\,,
    \end{cases}
\end{equation}
with $\omega>0$, $\epsilon\ll1$ the perturbative parameter, and $Y_{00}(\Omega)=1/\sqrt{4\pi}$. Up to the first nontrivial order in $\epsilon$, the mean number of particles created during the oscillation is
\begin{equation}
    \begin{aligned}
    \langle N\rangle
    &=
    \frac{\epsilon^2}{\pi}
    \sum_{\ell=0}^{\infty}(2\ell+1)
    \sum_{n=1}^{\infty}
    \sum_{k=1}^{\infty}
    \Bigg\{
    \sum_{\sigma=\pm}
    \left(v^{\sigma}_{\ell nk}\right)^2
    \frac{\sin^2\left[\frac{T}{2}\left(W_{\ell nk}-\sigma\omega\right)\right]}
    {\left(W_{\ell nk}-\sigma\omega\right)^2}
    \\
    &\hspace{3.2cm}
    +v^{-}_{\ell nk}v^{+}_{\ell nk}
    \frac{\cos(\omega T)}{W_{\ell nk}^2-\omega^2}
    \left[\cos(\omega T)-\cos(W_{\ell nk}T)\right]
    \Bigg\}\,,
\end{aligned}
\label{ape2}
\end{equation}
where
\begin{equation}
    W_{\ell nk}:=\omega_{\ell n}+\omega_{\ell k},
    \qquad
    \omega_{\ell n}=\frac{j_{\ell,n}}{R}\,,
\end{equation}
with $j_{\ell,n}$ the $n$-th zero of the spherical Bessel function of order $\ell$. The quantities $v^{\pm}_{\ell nk}$ are functions of the oscillation frequency $\omega$. See Ref.~\cite{sphere} for more details.

The case of interest is the limit $\omega R\gg 1$, together with long times, $T\gg 1/\omega$, since in the example of Sec.~\ref{examples} we considered undamped oscillations. This case corresponds to taking the continuum limit of Eq.~\eqref{ape2}, in which three different contributions can be identified. The first term, proportional to $(v^{+}_{\ell nk})^2$, contains a delta-function approximant, since, in the distributional sense,
\begin{equation}
    \frac{\sin^2\left[\frac{T}{2}(W_{\ell nk}-\omega)\right]}
    {(W_{\ell nk}-\omega)^2}
    \;\xrightarrow[T\to\infty]{}\;
    \frac{\pi T}{2}\,\delta(W_{\ell nk}-\omega)\,.
\end{equation}
The second term, proportional to $(v^{-}_{\ell nk})^2$, gives an analogous delta function evaluated at $W_{\ell nk}+\omega$, but this contribution vanishes because $W_{\ell nk}>0$ and $\omega>0$. Finally, the third term, involving $v^{-}_{\ell nk}v^{+}_{\ell nk}$, also has an approximate delta function centered at $W_{\ell nk}-\omega$:
\begin{equation}
  \frac{\cos(\omega T)-\cos(W_{\ell nk}T)}
  {W_{\ell nk}^{2}-\omega^{2}}
  \;\xrightarrow[T\to\infty]{\omega\approx W_{\ell nk}}\;
  \frac{\pi}{\omega}\sin(\omega T)\,\delta(W_{\ell nk}-\omega)\,.
\end{equation}
However, unlike the first term, this contribution is not proportional to the time scale $T$. It is therefore negligible, in the large-time limit, compared with the contribution that grows linearly with $T$. Thus, in this regime, the leading contribution comes only from the first term in Eq.~\eqref{ape2}.

To take the continuum limit in Eq.~\eqref{ape2}, we replace the sums by integrals according to
\begin{equation}
    \sum_{\ell=0}^{\infty}(2\ell+1)\mapsto\int_{0}^{\infty}\mathrm{d}\nu\,2\nu\,,
    \qquad\text{and}\qquad
    \sum_{n=1}^{\infty}\sum_{k=1}^{\infty}
    \mapsto
    \int_{0}^{\infty}\mathrm{d}\omega_{1}\frac{\partial n}{\partial\omega_{1}}
    \int_{0}^{\infty}\mathrm{d}\omega_{2}\frac{\partial k}{\partial\omega_{2}}\,,
\end{equation}
where $\omega_1=\omega_{\ell n}$, $\omega_2=\omega_{\ell k}$, and $\nu=\ell+1/2$. Using $W_{\ell nk}=\omega$ to evaluate the coefficients $v^{+}_{\ell nk}$ in Ref.~\cite{sphere}, we obtain $v^{+}_{\ell nk}=\frac{1}{2}\sqrt{\omega_{\ell n}\omega_{\ell k}}$, thus
\begin{equation}
    \langle N\rangle
    \approx
    \frac{1}{4}\epsilon^2 T
    \int_{0}^{\infty}\mathrm{d}\nu\,\nu
    \int_{0}^{\infty}\mathrm{d}\omega_{1}\frac{\partial n}{\partial\omega_{1}}
    \int_{0}^{\infty}\mathrm{d}\omega_{2}\frac{\partial k}{\partial\omega_{2}}\,
    \omega_{1}\omega_{2}\,
    \delta\!\left(\omega_1+\omega_2-\omega\right)\,.
\label{ape3}
\end{equation}
We therefore need to approximate the zeros of the spherical Bessel functions in order to compute the densities $\partial n/\partial\omega_{1}$ and $\partial k/\partial\omega_{2}$. The condition $\omega R\gg1$ implies that the relevant zeros have large radial quantum numbers. Moreover, because of the weight $2\nu$ in the angular integral, large values of $\nu$ also contribute. We therefore need the asymptotic distribution of zeros in the regime where both $n$ and $\nu$ are large. For this purpose, we use the Debye asymptotic expansion in the oscillatory region $x>\nu$ \cite{asymp},
\begin{equation}
    j_{\ell}(x)
    \sim
    \frac{1}{x}\left(1-\frac{\nu^2}{x^2}\right)^{-1/4}
    \cos\left(\xi(\nu,x)-\frac{\pi}{4}\right)\,,
    \qquad
    \xi(\nu,x)
    =
    \sqrt{x^2-\nu^2}
    -\nu\arccos\left(\frac{\nu}{x}\right)\,,
\end{equation}
so the $n$-th zero of the spherical Bessel function of order $\ell=\nu-1/2$ satisfies, in this regime, $\xi(\nu,j_{\ell,n})=n\pi$. Thus,
\begin{equation}
    \frac{\partial n}{\partial\omega_{1}}
    \approx
    \frac{R}{\pi}
    \sqrt{1-\frac{\nu^2}{R^2\omega_1{}^2}}\,,\qquad \omega_1R\geq\nu\,,
\end{equation}
and similarly for $\partial k/\partial\omega_{2}$. Therefore, after rescaling the integration variables to dimensionless ones, Eq.~\eqref{ape3} becomes
\begin{equation}
    \langle N\rangle
    \approx
    \frac{1}{2\pi^2}\epsilon^2 T \omega^5R^4
    \int_{0}^{1/2}\mathrm{d}x\,x(1-x)
    \int_{0}^{x}\mathrm{d}y\,y\,
    \sqrt{1-\frac{y^2}{x^2}}
    \sqrt{1-\frac{y^2}{(1-x)^2}}\,,
\label{ape4}
\end{equation}
which can be evaluated explicitly:
\begin{equation}
    \frac{\langle N\rangle}{T}
    \approx
    \frac{1}{720\pi^2}\epsilon^2\omega^5R^4\,.
\label{ape5}
\end{equation}
This agrees with the high-frequency limit of Eq.~\eqref{sphere_mean_number} for $\ell=0$.
%====================================================================
%====================================================================

\bibliographystyle{unsrtnat} % plain, alpha, etc.
\bibliography{bibliography} % without the .bib extension

\end{document}